\documentclass[fleqn,usenatbib]{mnras}

\usepackage{newtxtext,newtxmath}
\usepackage[T1]{fontenc}
\usepackage{booktabs}
\usepackage{array}
\usepackage{graphicx}

\DeclareRobustCommand{\VAN}[3]{#2}
\let\VANthebibliography\thebibliography
\def\thebibliography{\DeclareRobustCommand{\VAN}[3]{##3}\VANthebibliography}

\usepackage{graphicx}	% Including figure files
\usepackage{amsmath}	% Advanced maths commands
\usepackage{float}
\usepackage{orcidlink}
\usepackage{lineno}

\title[GW from accretion-induced starquakes]{Deciphering Accretion-Driven Starquakes in Recycled Millisecond Pulsars using Gravitational Waves}

\author[S. Chatterjee et al.]{
Sagnik Chatterjee \orcidlink{0000-0001-6367-7017} ,$^{1}$\thanks{sagnik18@iiserb.ac.in}
Kamal Krishna Nath \orcidlink{0000-0002-4657-8794},$^{2}$ \thanks{ kknath@niser.ac.in}
Ritam Mallick \orcidlink{0000-0003-2943-6388} $^{1}$\thanks{ mallick@iiserb.ac.in}
\\
$^{1}$Department of Physics, Indian Institute of Science Education and Research Bhopal, Madhya Pradesh, India, 462066 \\
$^{2}$School of Physical Sciences, National Institute of Science Education and Research,
An OCC of Homi Bhabha National Institute, Jatni-752050, India}
\date{Accepted XXX. Received YYY; in original form ZZZ}

\pubyear{2015}

\begin{document}

\label{firstpage}
\pagerange{\pageref{firstpage}--\pageref{lastpage}}
\maketitle

\begin{abstract}	
Recycled millisecond pulsars are susceptible to starquakes as they are continuously accreting matter from their binary companion. A starquake happens when the rotational frequency of the star crosses its breaking frequency. In this study, we perform a model analysis of an accreting neutron star suffering a starquake. We analyse two models: a spherical star with accreting mountains and a deformed star with accreting mountains. We find that as the star crosses the breaking frequency and suffers a starquake, there is a sudden change in the continuous gravitational wave signal arriving from it. The amplitude of the gravitational wave
signal increases suddenly both for the spherical and deformed star. For the spherical star, the accreting matter entirely dictates the amplitude of the gravitational wave. For the deformed star, both the accreting matter and the deformation from spherical symmetry play a significant role in determining the amplitude of the gravitational wave signal. This sudden change in the continuous gravitational wave signal in recycled millisecond pulsars can be a unique signature for such pulsars undergoing a starquake.

\end{abstract}


%%%%%%%%%%%%%%%%%%%%%%%%%%%%%%%%%%%%%%%%%%%%%%%%%%

%%%%%%%%%%%%%%%%% BODY OF PAPER %%%%%%%%%%%%%%%%%%
%\linenumbers
\section{Introduction}

The advent of modern gravitational wave (GW) detectors has opened a new horizon in the field of astrophysics, resulting in the detection of GW signals from binary black hole mergers \citep{Abbot2016} and, subsequently, the first-ever detection of signals from binary neutron star (BNS) mergers \citep{Abbot2017}. Developments of more state-of-the-art and sensitive detectors allow us to detect more such GW signals. So far, we have only been able to measure GW signals from mergers \citep{Bauswein, Hanauske, Haque}, but single neutron stars (NSs) can also be the source of continuous GW signals \citep{Lasky, Bejger}. Continuous GW emission \citep{Pagliaro_2023, Riles, Wette} from a single NS can arise due to the change in mass quadrupole moment of the star; this is possible where the stars are deformed, containing some mountains on its surface. Deformations of NSs can be a result of high angular velocity \citep{Bobur, Chatterjee} (pulsar) or very high magnetic field \citep{Lander_2017, Kuzur} (magnetars). The change in the mass quadrupole moment of such stars can result in the emission of constant GW signals \citep{Bonazzola}. \\

\indent The process by which matter from a companion star or an accretion disc falls on the surface of the primary star is known as accretion \citep{ Shakura, Wang2016, Wolf, Abolmasov, Fantina, Lyutikov, Largani}. High mass X-ray binaries (HMXBs) show various types of accretion \citep{kret}, like accretion from stellar winds \citep{karino}, Roche lobe overflow \citep{ducci, Mellah, Yu}. One of the first observable accreting X-ray sources was recorded in 1971 by NASA's Uhuru satellite. Centaurus X-3 \citep{uhuru_cent} and Hercules X-1 \citep{uhuru_hercules} were the first of such observed accreting X-ray binary systems. Accretion can give rise to interesting events like glitch \citep{Galloway, Serim, Yim2020, Yim2023}, spin-down \citep{Matsuda, Illarionov, Prasad22}, and spin-up \citep{Bhattacharyya_2017} in NS. \\

Millisecond pulsars are thought to be NSs, which are spun up by accretion from their binary companion \citep{Riggio_2008, Papitto2013, L_Wang, disalvo2020accretion}. They are detected as radio pulsars \citep{Campana_2004, Hessels, Iacolina}, X-ray bursts \citep{Strohmayer_2017, Sanna, Bult_2018} or even in gamma rays \citep{GRENIER2015641, Wilhelmi2016, Chang}. They are also called recycled pulsars as they are old pulsars whose electromagnetic phenomena have been recycled by mass accretion from their binary companion \citep{Wijnands1998, Li_2021}. The standard evolution model fails to account for all millisecond pulsars, as their younger counterpart have larger magnetic fields. Therefore, it is believed that at least there are two different types of millisecond pulsars \citep{Lorimer1998-lm}. One interesting property that an accretion-induced millisecond pulsar shows is the starquakes.  \\

\indent Starquakes are events occurring on the surface of the NS \citep{Bransgrove_2020, Lu} where the crust of the star starts to crack \citep{Link, Kerin} under the development of stress \citep{Franco, Giliberti_strain}. This leads to the rearrangement of mass on the star's surface, resulting in a change in the ellipticity of the star \citep{Giliberti_2022}. The increasing stress upon the crust once it reaches a critical threshold value \citep{Christensen} triggers a quake, which can also result in oscillations \citep{Keer}.\\

\indent In this work, we consider that the star is composed of a fluid core with an elastic crust \citep{Giliberti}. The star has two mountains at its magnetic poles \citep{Sousa}. We consider two models, one having a spherically symmetric star with mountains and another a deformed star consisting of mountains. The presence of the mountains results in the generation of continuous GW signal from the star \citep{Haskell}. The emission of the GW signal can be calculated from the change in the mass quadrupole moment \citep{Bonazzola} of the NS. As the star accumulates of mass on the its surface, the stress on the star surface increases and ultimately the stars suffers a starquake. The mass quadrupole moment changes abruptly from the pre-starquake star, changing the GW amplitude abruptly. We compare these GW signals to study a starquake scenario in a star leading to a change in the pre-existing GW signals. The work studies the possibility of such signals to be detected by present and future detectors. \\

\indent This paper is organised in the following way: in section \ref{model}, we have introduced our model describing the structure of the crust and core of the star. We also introduce the orientation of the mountains and the NS model for misalignment. Section \ref{results} details the results we obtained and the salient features of our work. The conclusion and summary of this work can be found in section \ref{summary}.

%------------------------------------------------------------------------------------------
% ======================================MODEL==============================================
%-------------------------------------------------------------------------------------------

\section{Model} \label{model}
 \begin{figure}
	\includegraphics[scale=0.32]{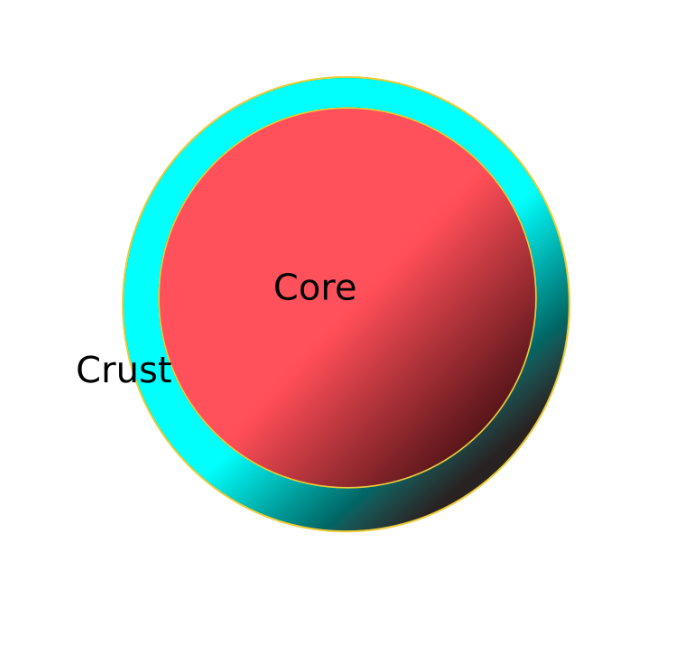}
	
	\caption{The schematic diagram of our model indicating the core and crust of the star. The red-coloured portion shown in the figure represents the star's core, and the cyan-coloured portion indicates the crust.}
	\label{core}
\end{figure}

We begin the analysis by considering a stellar configuration in which the star comprises two distinct components: the core and the crust, as shown in Fig \ref{core}. 
In this analysis, we consider the Franco-Link-Epstein or FLE model, which is a Newtonian model of an incompressible star \citep{Franco}. The model assumes the core to be fluid in nature \citep{Giliberti_2022} whereas the crust is rigid and can break upon sufficient stress. Once the breaking threshold stress is reached the brittle crust cracks. The breaking threshold is reached first at the crust-core interface near the equatorial plane. Under the process of accretion, the infalling matter tries to spin up the star, as a result of which the stress on the crust increases, thus resulting in a change in ellipticity, resulting ultimately in a starquake. 

%------------------     Breaking Frequency-----------------------------------------

 \subsection{Breaking Frequency} \label{breaking frequency}

The maximum amount of strain the crust can hold before it breaks down at a specific frequency (called the breaking frequency), resulting in a starquake, must be estimated. In the present calculations, we use the FLE model (which is discussed in Appendix \ref{FLE}), having a discontinuous shear modulus at the interface of the crust and core of the star \citep{Franco}.. For a realistic equation of state (EoS), the FLE model is used to find the strain angle of the star. The equation of state (EoS) used here is {\ttfamily DD-ME2BPS}, which is constructed by taking {\ttfamily DD-ME2} (\cite{Lalazissis}) as the hadronic part of EoS and considering {\ttfamily BPS} \citep{Baym, Negele} EoS as the crust. The EoS follows the current astrophysical constraints giving a maximum mass of $ M_{TOV} = 2.48 M_{\odot}$ satisfying the maximum mass limit \citep{Antoniadis, Cromartie, Fonseca} and having a binary tidal deformability $\tilde \Lambda \leq 800$ \citep{Abbot2017, Raithel_2018}. \\

 We first define a position vector $\vec{x}$ of the matter in the star's initial configuration of the star \citep{Giliberti}. Under rotation, this point shifts to a new point, namely $\vec{r}(x)$. This can be defined as :
\begin{align}
    \vec{r}(x)=\vec{x}+\vec{u}(x)
\end{align}
Here, $\vec{u}$ is called the displacement vector and is a function of the radius and the azimuthal angle, $u(r,\theta)$. We define a strain tensor as \citep{Love} :
\begin{align}\label{strain}
    u_{ij}=\frac{1}{2}\left(\frac{\partial u_{i}}{\partial x_{j}}+\frac{\partial u_{j}}{\partial x_{i}}\right)
\end{align}
The strain angle $\alpha$ can be found by the difference between the local maximum and local minimum between the eigenvalues of eq \eqref{strain} \citep{landau}. \\

To connect the model with a realistic scenario, from here onwards, we consider our model neutron star to have physical characteristics similar to that of Cyg-X2. Cyg-X2 has one of the highest mass accretion rates, and therefore, it suits best for our analysis \citep{Ozel}. The mass of Cyg-X2 is approximately $1.43\pm 0.10 M_{\odot}$ \citep{King} and the rotational frequency is $346\pm29 \text{Hz}$ \citep{Wijnands}. As there is no radius measurement of Cyg-X2, hence the values of the radii's used in our calculations are chosen in alliance with the radius estimate of PSR J0030+0451 \citep{Miller_2019} (radius estimate of $R = 13.02 \pm ^{1.24}_{1.06} \text{km}$ for a mass of $M = 1.44\pm^{0.15}_{0.14} M_{\odot}$).  
In the subsequent sections, we introduce two models for our analysis and the values of the variables considered for the models are shown in Table \ref{table1}. The choice of misalignment angle and line of sight angle are not unique and are chosen for convenience of our calculations. Other choices would not change the quantitative analysis performed in subsequent sections.

\begin{table}
     \centering
     \begin{tabular}{c c c}
     \hline
          & \textbf{\textit{model I}} & \textbf{\textit{model II}} \\ \hline  \\
         Mass [$M_{\odot}$] & $1.46$ &   $1.46$  \\  \\ \hline \\
         Initial angular frequency [Hz]&  $346$ & $346$ \\ \\  \hline \\
          Equatorial radius [km] &  $12.3$ & $13.48$ \\ \\  \hline \\
          Polar radius [km] & $12.3$ & $10.78$ \\ \\ \hline \\
          $\delta M$ [$M_{\odot}$] & $10^{-4}$ &   $10^{-4}$  \\  \\ \hline \\
          
          Misalignment angle ($\chi$) & $\pi / 6$ &   $\pi/6$  \\  \\ \hline \\
          Line of sight angle ($i$) & $\pi / 9$ &   $\pi / 9$  \\  \\ \hline \\

     \end{tabular}
     \caption{ The Table shows the initial values of the variables we have chosen for each model.}
     \label{table1}
 \end{table}

\begin{figure}
    \includegraphics[scale=0.60]{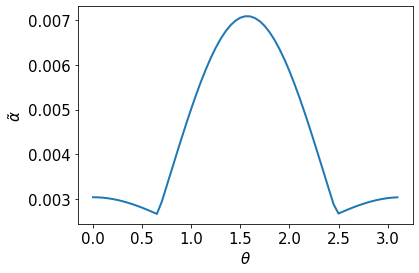}
	\caption{The plot of the strain angle $\Tilde{\alpha}$ against the azimuthal angle $\theta$.}
	\label{FLE_Strain2}
\end{figure}

The radius of the core extends up to $0.95\:R$ (where $R$ denotes the star's radius) from the star's centre, and the outer layer is the crust \citep{Franco, Giliberti}. To calculate the breaking frequency, we use the Tresca criterion \citep{Christensen}, which gives:
\begin{align} \label{tresca1}
    \alpha=\frac{\sigma_{max}}{2}
\end{align}
where $\sigma_{max}$ is called the breaking strain \citep{Christensen}.
We redefine the strain angle \citep{Giliberti_2022} as:
\begin{align}\label{tresca2}
    \alpha=\Tilde{\alpha} \nu^{2}
\end{align}
where $\tilde {\alpha}$ is a term that depends on the structure of the NS. The plot of $\tilde {\alpha}$ against the azimuthal angle $\theta$ is shown in Fig \ref{FLE_Strain2}.
The value of the maximum strain angle ($\tilde {\alpha}_{max}$) is near the equator of the star. Using this plot, we calculate the breaking frequency of the star. Substituting eq \eqref{tresca1} in eq \eqref{tresca2}, we get the equation of the breaking frequency of the NS satisfying the Tresca criterion as
\begin{align} \label{breaking_freq}
    \nu_{b}=\sqrt{\frac{\sigma_{max}}{2\Tilde{\alpha}_{max}}}
\end{align}
%There is no clear agreement over the value of $\sigma_{max}$. 
The values of $\sigma_{max}$ can range from $10^{-5}$ as given by \cite{Ruderman} to $10^{-1}$ as suggested by \cite{Horowitz}. However, in our calculations of breaking frequency, we take the value of $\sigma_{max}$ to be 0.04 \citep{Baiko}. This gives the value of our breaking frequency to be $\nu_{b}=580 \: \textup{Hz}$. It is to be noted that the breaking frequency is also dependent on the mass of the star (shown in Appendix \ref{FLE}). 
%The possibility of attaining such a frequency is discussed in Appendix \ref{Time_required}.
%=======================================================================================
%=============================Moment Of Inertia==========================================
%======================================================================================
\subsection{Moment of Inertia Calculations For A Spherical Star : \textit{model I}} \label{model1}
\begin{figure}
    \includegraphics[scale=0.32]{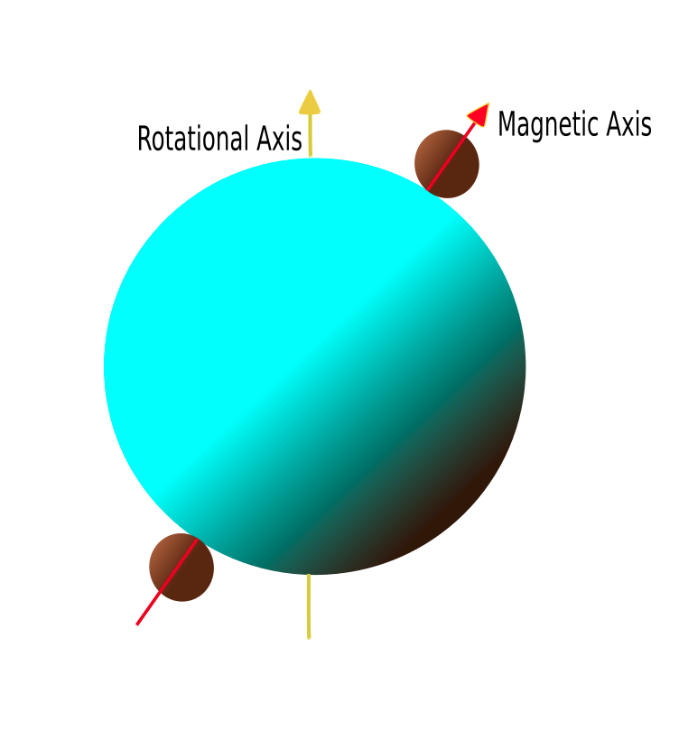}

	\caption{\textit{Model I:} The schematic diagram of our model of the star. The cyan-colored sphere denotes the pulsar. The two spherical balls on the star's surface denote the mountains. The red colored line denotes the magnetic axis, and the yellow-colored line denotes the rotational axis.}
	\label{schematic}
\end{figure}

Fig \ref{schematic} shows the schematic diagram of the model. We consider a spherically symmetric star with two spherical mountains present at the two magnetic axes, and we assume that there is no deformation due to the star's rotation on its axis. In real scenarios, we do not expect the mountains to be spherical; however, considering them to be spherical simplifies the calculations. Since we consider the radii of the mountains to be negligible compared to the radius of the star (in the subsequent section), the mountains' shape does not substantially affect our calculations. The mass of the mountains is the more determining factor. The mountains occur at the magnetic poles. This is due to the fact that the presence of a strong magnetic field channels the in-falling matter along the magnetic field lines; as a result, more matter is concentrated at the magnetic poles of the star \citep{Longhair, Koldoba_2002}. The moment of inertia components of the spheroid (without the presence of any mountains or deformation of the star) along the magnetic axis are given by:
\begin{equation}
    I =  \begin{pmatrix}
  \dfrac{2}{5}MR^{2} & 0 & 0\\ 
  0 & \dfrac{2}{5}MR^{2} & 0\\
  0 & 0 & \dfrac{2}{5}MR^{2}
\end{pmatrix}
\end{equation}
where $R$ is the radius and $M$ is the mass of the star. When the mass is accreted on the magnetic poles of the NS, the moment of inertia changes. The new moment of inertia, taking into consideration the mountains, takes the form:
\begin{equation} \label{Inew}
    I' =  \begin{pmatrix}
  \dfrac{2}{5}MR^{2} + 2 \delta m R^{2} & 0 & 0\\ 
  0 & \dfrac{2}{5}MR^{2} + 2 \delta m R^{2} & 0\\
  0 & 0 & \dfrac{2}{5}MR^{2}+ 2 \dfrac{2}{5} \delta m a^{2}
\end{pmatrix}
\end{equation}
Where $\delta m$ represents the total mass accreted by the star and $a$ is the radius of the spherical globules of accreted mass (with the condition $a<<R$ \citep{Sousa} ). The change in mass quadrupole moment \citep{Bonazzola} is given as:
\begin{equation}\label{mass_quad1}
    \mathcal{I}_{ij}=-I_{ij} + \frac{1}{3}I_{k}^{k} \delta_{ij}
\end{equation}
Evaluating eq \eqref{mass_quad1} gives the mass quadrupole moment by using the moment of inertia obtained from eq \eqref{Inew} as  :
\begin{equation}
    \mathcal{I} = \begin{pmatrix}
        -\dfrac{2}{3} \delta m R^{2} & 0 & 0\\
        0 & -\dfrac{2}{3} \delta m R^{2} & 0\\
        0 & 0 & \dfrac{4}{3} \delta m R^{2}\\
    \end{pmatrix}
\end{equation}

The mass quadrupole moment has been calculated along the magnetic axis of the NS. We perform two transformations to go to the rotational axis. The first transformation accounts for the star's rotation, and the second rotates our frame along the inclination axis of the star. The resulting transformation matrix is given as:
\begin{equation} \label{transform}
    P=  \begin{pmatrix}
  \cos \Omega t & - \sin \Omega t & 0\\ 
  \sin \Omega t & \cos \Omega t & 0\\
  0 & 0 & 1
\end{pmatrix} 
\times
  \begin{pmatrix}
 1 & 0 & 0\\ 
  0 & \cos \chi & \sin \chi\\
  0 & -\sin \chi & \cos \chi\\
\end{pmatrix} 
\end{equation}
Here, $\Omega$ is the angular velocity of the star, and $\chi$ is the inclination angle of the star. The mass quadrupole moment ($\mathcal{I}^{dist}$) in the frame of the rotation axis is given by, 
\begin{align}
    \mathcal{I}^{dist}=P \times \mathcal{I} \times P^{t}
\end{align} 
here $P^{t}$ is the transpose of the P-matrix from eq \eqref{transform}. The leading term in the gravitational radiation field is given as \citep{Bonazzola, Bejger}: 
\begin{equation} \label{GW_lead}
    h_{ij}^{TT} = \frac{2G}{c^{4}} \frac{1}{r} \left[P_{i}^{k}P_{j}^{l}-\frac{1}{2}P_{ij}P^{kl}\right] \Ddot{\mathcal{I}}_{kl}
\end{equation}
where $G$ is the gravitational constant, $c$ being the speed of light, $r$ being the distance of the source from the observer and $P_{ij}$ represents the transverse projection operator and is given as $P_{ij}=\delta_{ij} - n_{i}n_{j}$. Putting the value of $\mathcal{I}^{dist}$ in eq \eqref{GW_lead} we get the following:
\begin{align}\label{hplus_form}
    & h_{+}=h_{0} \sin \chi \left[\frac{1}{2}\cos \chi \sin i \cos i \cos \Omega t - \sin \chi \frac{1+ \cos^{2}i}{2} \cos 2\Omega t\right] \nonumber \\
    & h_{\times}=h_{0} \sin \chi \left[\frac{1}{2} \cos \chi \sin i \sin \Omega t - \sin \chi \cos i \sin 2\Omega t\right]
\end{align}
where  $i$ denotes the line of sight angle and
\begin{align} \label{ho}
    &h_{0}:= - \frac{6G}{c^{4}} \mathcal{I}_{zz}^{dist} \frac{\Omega^{2}}{r}\nonumber \\
    &\mathcal{I}_{zz}^{dist}=  \frac{4}{3} \delta m R^{2}
\end{align}

%=======================================================================================
%=======================Moment Of Inertia For Deformed Star==========================================
%======================================================================================

\subsection{Moment of Inertia Calculations For Deformed Star : \textit{model II}} \label{model2}
\begin{figure}
	\includegraphics[scale=0.38]{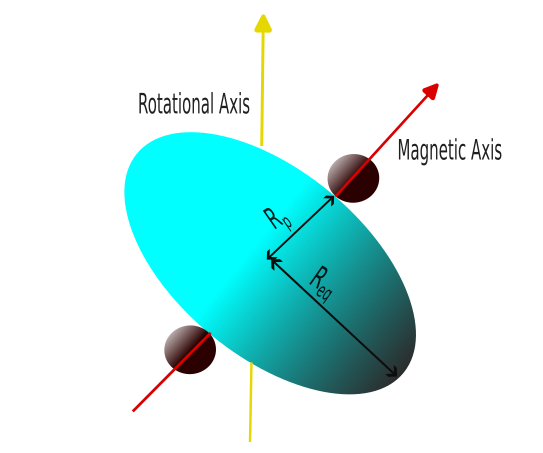}
	
	\caption{\textit{Model II: }The schematic diagram showing a cyan-colored deformed star about its magnetic axis. The brown balls represent spherical mountains at their magnetic poles.} The red-colored line denotes the magnetic axis, and the yellow-colored line denotes the rotational axis.
	\label{schematic_Deformed}
\end{figure}
Considering the star to be spherically symmetric is an ideal scenario. In real scenarios, the star is not symmetric and can get deformed either due to rotation or due to the presence of strong magnetic fields. In this model, we consider the star to be an oblate spheroid having mountains at the magnetic axis, as shown in Fig \ref{schematic_Deformed}. We consider that the star is rotating with an angular frequency of $346$ Hz (similar to that of Cyg-X2) and has a surface magnetic field of $\sim$ $10^{15}$ G. For the rotation, the ratio of $\frac{R_p}{R_{eq}}$ comes out to be $\approx$ $0.96$ which is very close to $1$. Where $R_{eq}$ is the equatorial radius, $R_{p}$ is the polar radius of the star. For a perfectly spherical star (which is not deformed), the radius at the poles and the equator becomes the same (i.e., $R_{eq} = R_{p}$). The ratio of $\frac{R_p}{R_{eq}}$ close to 1 means that the shape of the star does not change significantly due to rotation. However, for a magnetic field of $\sim$ $10^{15}$ G the ratio of $\frac{R_p}{R_{eq}}$ come out to be $\approx$ $0.80$. It is a significant deformation from the spherical symmetry of the star.

We used RNS code (\cite{1998Nozawa,1995Stergioulas,1994Cook,1989Komatsu}) for modelling the rotating NS and XNS code (\cite{2011Bucciantini,2007Del,2014Pili}) for modelling the magnetized counterpart. Therefore, we have approximated that the star's rotation does not contribute to any significant deformations around the rotational axis, and the reason for the oblateness around the magnetic axis is due to a strong magnetic field.

The moment of inertia components of an oblate spheroid (without the presence of any mountains) in the magnetic axis frame is given by:
\begin{equation}
	I =  \begin{pmatrix}
		\dfrac{M}{5}(R_{eq}^{2}+R_{p}^{2}) & 0 & 0\\ 
		0 & \dfrac{M}{5}(R_{eq}^{2}+R_{p}^{2}) & 0\\
		0 & 0 & \dfrac{2}{5}M R_{eq}^{2}
	\end{pmatrix}
\end{equation}
 The accretion of masses at the magnetic poles will result in a change in the moment of inertia. The new moment of inertia takes the form:
\begin{equation}  \label{Iprime_Deform}
	I' =  \begin{pmatrix}
		\dfrac{M}{5}(R_{eq}^{2}+R_{p}^{2})  & 0 & 0\\ 
  		+ 2 \delta m R_{p}^{2} & & \\ \\

		0 & \dfrac{M}{5}(R_{eq}^{2}+R_{p}^{2})  & 0\\
  		 &  + 2 \delta m R_{p}^{2} & \\ \\

		0 & 0 & \dfrac{2}{5}M R_{eq}^{2} \\
  		 &  & + 2 \dfrac{2}{5} \delta m a^{2}

	\end{pmatrix}
\end{equation}
Following a similar prescription as that of the previous model, the GW amplitude is given as
\begin{align}
	& h_{+}=h_{0} \sin \chi \left[\frac{1}{2}\cos \chi \sin i \cos i \cos \Omega t - \sin \chi \frac{1+ \cos^{2}i}{2} \cos 2\Omega t\right] \nonumber \\
	& h_{\times}=h_{0} \sin \chi \left[\frac{1}{2} \cos \chi \sin i \sin \Omega t - \sin \chi \cos i \sin 2\Omega t\right]
\end{align}\label{hplus_deform}
where 
\begin{align}\label{h0_modelII}
	&h_{0}:= - \frac{6G}{c^{4}} \mathcal{I}_{zz}^{dist} \frac{\Omega^{2}}{r}\nonumber \\
	&\mathcal{I}_{zz}^{dist}=  \frac{2}{15} \left(-10 \: \delta m \: R_{p}^2+M \left( R_{eq}^2 - R_p^{2}\right)\right)
\end{align}

%=======================================================================================
%=========================Change In Ellipticity due to Starquake========================
%=======================================================================================
\subsection{Change in ellipticity due to starquake} \label{MomentI}
One of the possible consequences of starquakes is the change in the ellipticity of the star. As a starquake occurs, the mass accumulated on the star's surface tries to reallocate themselves on the star's surface. This results in the change of ellipticity of the star. In this subsection, we calculate the ellipticity change due to a starquake.\\
\indent Initially, the amount of ellipticity of the configuration is given \citep{Kuzur} as:
\begin{align} \label{ellip}
    \epsilon_{i}=\Bigg| \frac{I_{zz}-I_{xx}}{I_{0}}\Bigg |
\end{align}
where $I_{zz}$ and $I_{xx}$ are the principal moment of inertia along the x-axis and the z-axis. The z-axis lies along the magnetic axis, and the x-axis and y-axis are on the star's equatorial plane. $I_{0}$ represents the moment of inertia of a non-deformed star. For this case, eq \eqref{ellip} takes the form:
\begin{align}
    \epsilon_{i}=\frac{2 \delta m R^{2}}{I_{0}}
\end{align}

For an accreted mass of about $\delta m = 10^{-4} \textup{M}_{\odot}$ we get the value as $\epsilon_{i}=3.43 \times 10^{-4}$ (for \textit{model I}). 
%In this case, we take the mass of the star to be $1.46 M_{\odot}$ similar to Cyg-X2 and the radius to be $12.3 \: \text{km}$.\\

With the increase in rotational velocity, the centrifugal force changes, thus changing the shape of the star. Considering these changes, the maximum ellipticity can be calculated by taking the difference between a pre-starquake model and the model after a starquake, as shown by \cite{Giliberti_2022}. In the pre-starquake model, the star is assumed to have an elastic crust and fluid core, and after the starquake, the star is only made of fluid. Thus, the difference in these two configurations allows us to estimate the maximum ellipticity that can occur due to the breakdown of the crust.\\
\indent The maximum change in ellipticity that occurs due to the starquake scenario is given as \citep{Giliberti_2022}:
\begin{align} \label{eccentricity}
    \epsilon_{max}=\frac{R^{3}}{3 I_{0}G}\left[\Phi^{F}(R)-\Phi^{E}(R)\right]
\end{align}
where $\Phi^{F}$ and $\Phi^{E}$ denote the total perturbed gravitational potential in spherical harmonics for a star with fluid configuration and a star with elastic crust configuration, respectively.\\

\indent The total perturbed gravitational potential in spherical harmonics can be defined as:
\begin{align} \label{grav_pot}
    \Phi(R)=-\frac{4 \pi G}{5 R^{3}} \int_{0}^{R} \: \rho(r) r^{4} \:dr
\end{align}
where $G$ is the gravitational constant, and $\rho$ is the energy density. Solving them we get the value of $\epsilon_{max}$ from eq \eqref{eccentricity} as $\epsilon_{max} \approx 2.25 \times 10^{-3}$.

%=======================================================================================
%====================================Results==========================================
%======================================================================================
\section{Results} \label{results}

To find the GW before and after the starquake for both \textit{model I} and \textit{model II}, consider the distance of observation to be $10 \: \text{kPc}$. We assumed the misalignment angle ($\chi$) of about $\pi /6$ and the inclination to the line of sight angle ($i$) to be $\pi / 9$. We take the initial angular frequency $346 \: \text{Hz}$ equivalent to the frequency of Cyg-X2 \citep{Wijnands,King}. Accretion will result in a gradual spin-up of the star, which, upon reaching the breaking frequency, will trigger a starquake, resulting in a change in the existing GW signal.\\

\begin{figure*} 
    \includegraphics[scale=0.35]{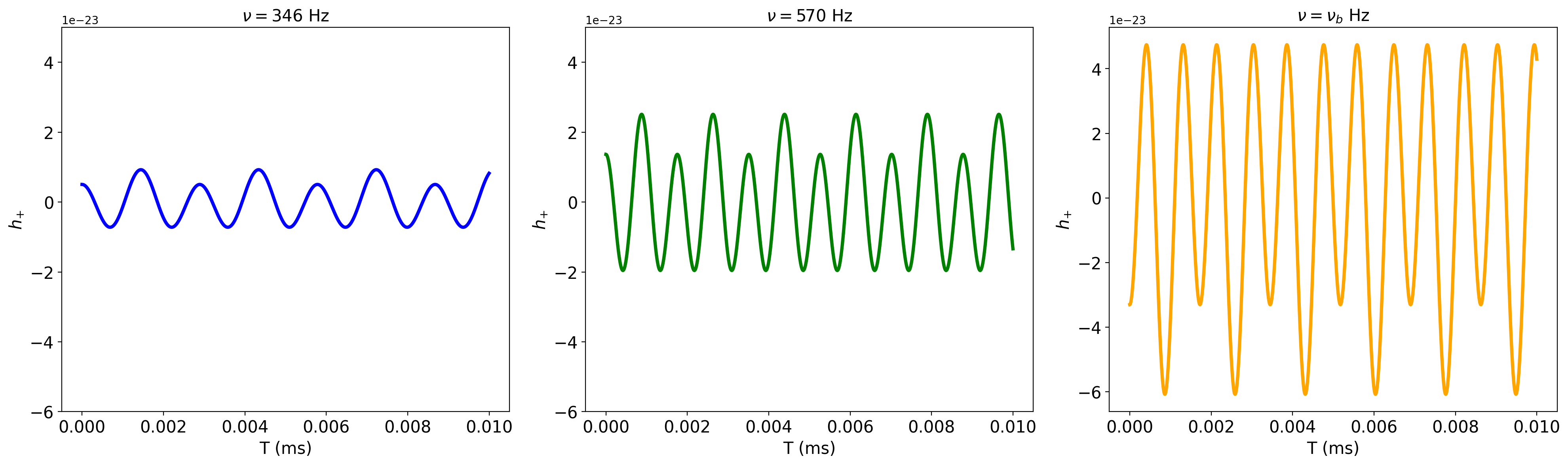}
	\caption{The plot of GW amplitude against time for a spherical star with magnetic mountain. The blue colored line denotes the GW at $\nu = 346 \text{Hz}$. The green coloured line shows the GW at $\nu = 570 \text{Hz}$ just before the starquake; the increase in amplitude is due to the star's accretion. The orange colored lines show how the GW would look after \textit{model I} has attained breaking frequency. }
	\label{acc}
\end{figure*}

Fig \ref{acc} shows the change in GW amplitude before and after the starquake for the \textit{model I}. At very low frequency (early stage of mass-accretion), the GW amplitude was $\sim 10^{-24}$. However, as the star ages, its frequency increases due to accretion; it tends towards the breaking frequency ($\nu_{b} = 580 \text{Hz}$). Till the starquake, the increase in frequency is slow and continuous; however, after the starquake, there is a sudden but substantial increase in the frequency ($\sim 10^{-23}$) (shown in Fig \ref{acc}). As more matter falls on the star's magnetic poles, the crust's stress keeps increasing. Once the breaking frequency is reached, there is a starquake, which results in a sudden change in the ellipticity of the star. This leads to a more significant change in quadrupole moment and, subsequently, to the GW amplitude.\\

For the case where the initial star is an oblate spheroid, we consider a stable star with $R_{eq} = 13.48 \: \text{km}$ and $R_{p} = 10.784 \: \text{km}$ having a mass of $1.46 \: M_{\odot}$. The choice of the equatorial radius lie within the radius range of PSR J0030+0451, as discussed previously. 

\begin{figure*}
	\includegraphics[scale=0.35]{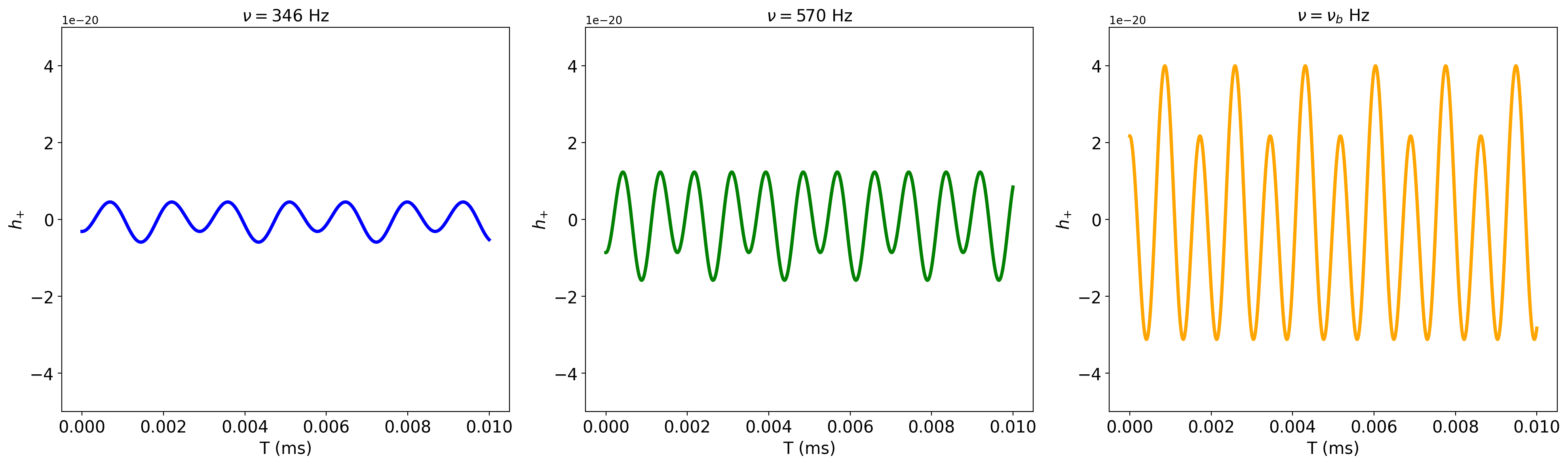}
	\caption{The plot of GW amplitude against time for a deformed star in the shape of an oblate spheroid with magnetic mountain. The blue colored line denotes the GW at $\nu = 346 \text{Hz}$. The green coloured line shows the GW at $\nu = 570 \text{Hz}$ just before the starquake; the increase in amplitude is due to the star's accretion. The orange colored line shows how the GW would look after \textit{model II} has attained breaking frequency. }
	\label{GW_dm}
\end{figure*}

Fig \ref{GW_dm} shows the variation of GW for an initial frequency of $\nu = 346 \text{Hz}$. Till the breaking frequency, the change in GW signal is slow and continuous; however, once the breaking frequency is breached, there is again a sudden but substantial change in the GW amplitude. After attaining the breaking frequency and the triggering of starquake, there is an increase in the amplitude of the signal from $\sim 10^{-20}$ to $\sim 10^{-19}$. Similar to the previous case, as the \textit{model II} star spins up, it reaches the breaking frequency, which triggers a starquake. There is a sudden and abrupt increase in the ellipticity after the starquake, increasing GW amplitude similar to the previous model. Also, in both cases, the frequency of the GW changes and therefore, there is a phase shift in the frequency due to starquake. 
 
%========================================================================================================
%=======================================================================================================

% ---------------- Prospects of detection --------------------------------
\subsection{Prospects of detection of starquakes from the GW signal}

\begin{figure*}
	\includegraphics[scale=0.55]{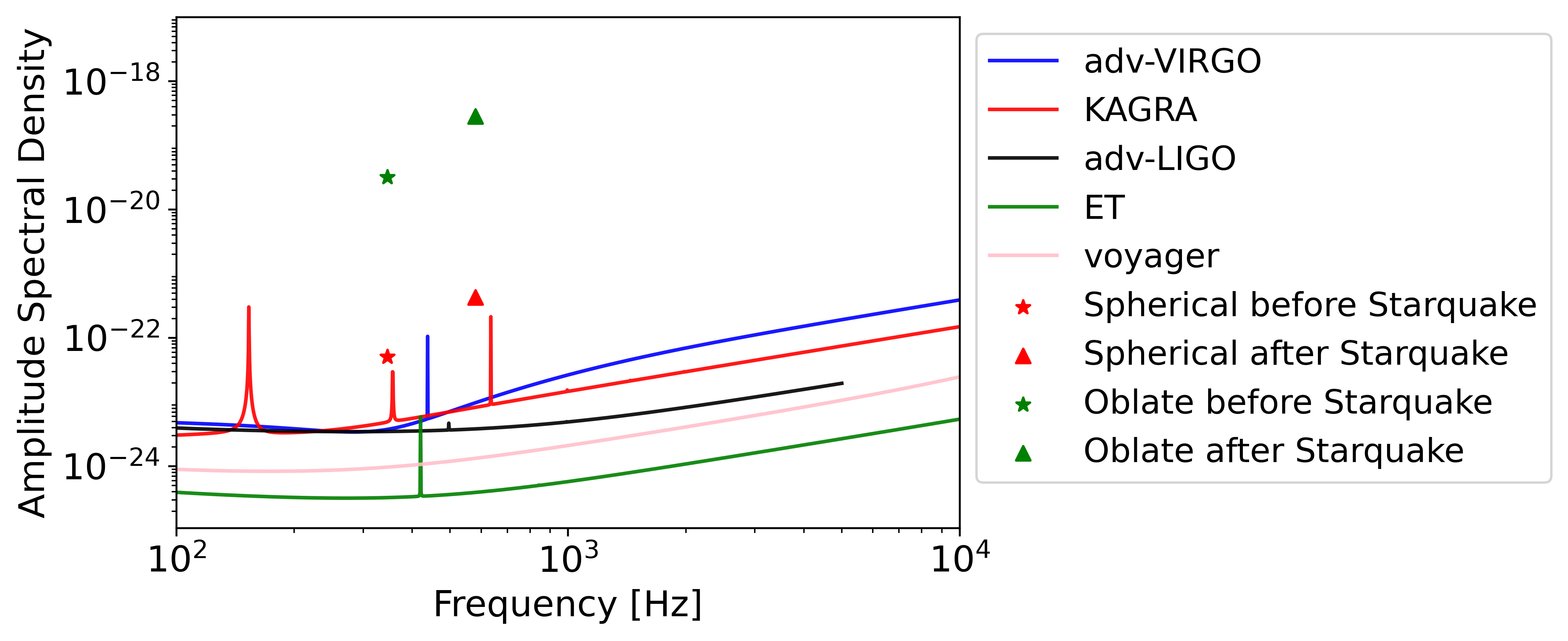}
	\caption{The plot of ASD against the frequency for our two models. The points show the $2 \tilde {h} \sqrt{f}$ for the two cases before and after the starquake. \textit{Model I} is represented by the red colored points, whereas \textit{model II} is represented by the green colored points.}
	\label{Starquake_Plots}
\end{figure*}

\indent With the advent of advanced GW detectors LIGO-VIRGO-KAGRA, any change in the GW signal may be detected. Present findings demonstrate that starquakes can be identified if a star's GW amplitude suddenly changes while the star's  GW signals are continuously monitored. \\
We next calculate the power spectral density, defined as \citep{Takami} :
\begin{align}
	\tilde{h}(f) = \sqrt{\dfrac{|\tilde{h}_{+}(f)|^{2} + |\tilde{h}_{\times}(f)|^{2}}{2}}
\end{align}
with 
\begin{align}
	& \tilde{h}_{+, \times} (f) = \int h_{+, \times}(t) \exp^{-i 2 \pi f t} \: dt \: \: , \text{where} \: \: f \geq 0
\end{align}
The value of $\Tilde{h}(f)$ is plotted against the frequency in Fig \ref{Starquake_Plots}. It shows that detecting such GW is possible as it lies within our current detector sensitivities. However, it is also to be noted that the current values depend heavily on various parameters one chooses, which are discussed in the subsequent sections.\\

Both the \textit{model I} and \textit{model II}, under current assumptions of the parameters, show an exciting scenario; before the occurrence of the starquake, the continuous GW signals lie in the current detector sensitivity ranges. For \textit{model I} and \textit{model II}, there is an increase in the amplitude spectral density (ASD) after a starquake.

%\newpage

% - ------------ Population of accreting pulsars ---------------------------

%\subsection{Population of accreting pulsars}

\begin{figure}
	\includegraphics[scale=0.60]{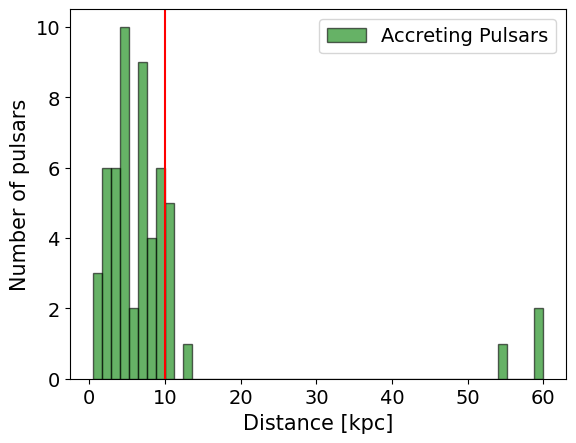}
	\caption{The histogram plot showing the number of accreting pulsars against the distance from the galactic plane. The red line represents the distance of 10 kpc.}
	\label{accretion}
\end{figure}

Next-generation detectors are expected to detect continuous GW signals from nearby pulsars. Therefore, it would be interesting to check how likely we are to observe a star with an accreting pulsar within a distance of 10 kpc (the distance of our source in our calculation). We plot a histogram for the number of accreting pulsars as shown in Fig \ref{accretion} \citep{Vall}. The number of pulsars inside the region of 10 kpc comes out to be 55.

%------------------ Effects of change in the position of mountain --------

\subsection{Effects of the position of magnetic mountain} \label{mountain_pos}
\begin{figure}
	\includegraphics[scale=0.30]{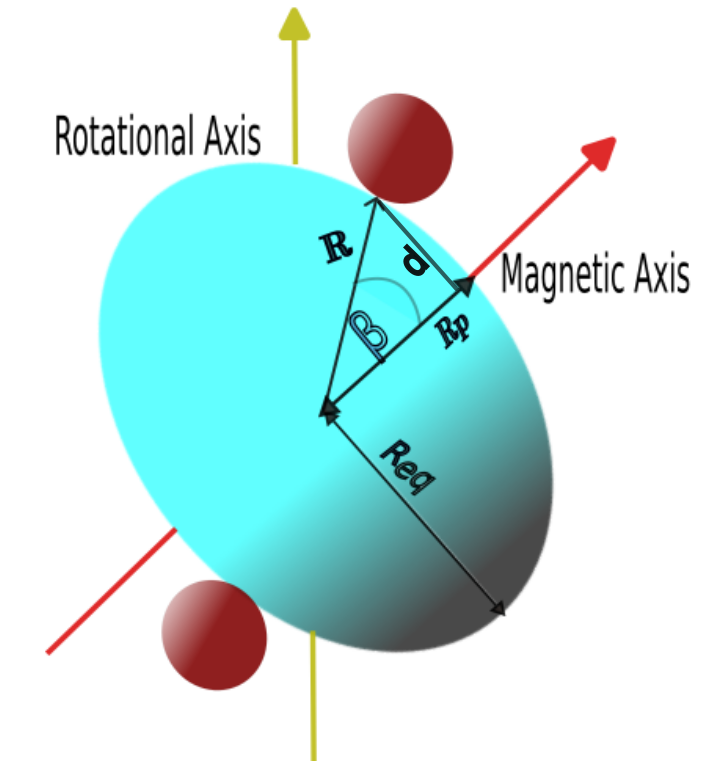}
	\caption{The schematic diagram is similar to Fig \ref{schematic_Deformed}; however, here, the position of the mountain is not at the magnetic pole.}
	\label{deform_R}
\end{figure}

It is interesting to check the dependence of the position of the mountains in our model on the GWs. To analyse this, we first take \textit{model II} and consider that the mountains are shifted from the magnetic pole at an angle $\beta$ as shown in Fig \ref{deform_R}.

We define the eccentricity of our oblate spheroid as:
\begin{align} \label{ecc}
    e= \sqrt{1- \dfrac{R_{p}^{2}}{R_{eq}^{2}}}
\end{align}
The position of the mountain is at a distance of $R$ from the centre of the star lying at an angle $\beta$ from the magnetic axis. The value of $R$ and perpendicular distance $d$ can be given as :
\begin{align}
    & R (\beta) = \dfrac{R_{p}}{\sqrt{1-e^{2}}} \times \sqrt{1-e^{2} \cos^{2} \beta}  \nonumber \\
    & d = R(\beta) \: \sin \beta
\end{align}
Eq \eqref{Iprime_Deform} in section \ref{model2} can be redefined as :

\begin{equation}  \label{Iprime_NewDeform}
	I' =  \begin{pmatrix}
		\dfrac{M}{5}(R_{eq}^{2}+R_{p}^{2})  & 0 & 0\\  
  		+ 2 \delta m R(\beta)^{2} &  & \\ 

		0 &\dfrac{M}{5}(R_{eq}^{2}+R_{p}^{2})  & 0\\
		   & + 2 \delta m R(\beta)^{2} &  \\

		0 & 0 & \dfrac{2}{5}M R_{eq}^{2} + 2 \delta m d^{2}\\
  		% &  & + 2 \delta m d^{2}

	\end{pmatrix}
\end{equation}

Calculating the mass quadrupole moment, we find :
\begin{align} \label{new_Idist}
     \mathcal{I}_{zz}^{dist} = \frac{2}{15} \Bigg[M(R_{eq}^{2} - R_{p}^{2}) \: - \: & 10 \delta m (R(\beta)^{2} - d^{2})  \Bigg]
\end{align}

Eq (\ref{new_Idist}) can be further simplified as :

\begin{align} \label{new_Idist2}
     \mathcal{I}_{zz}^{dist} = \frac{2}{15} \Bigg[M(R_{eq}^{2}  - R_{p}^{2}) \: -  \: &10 \delta m R(\beta)^{2} (1- \sin \beta^{2})\Bigg] 
\end{align}

The amplitude of our GW $h_{0} \propto \mathcal{I}_{zz}^{dist}$ which implies that the GW amplitude for \textit{model II} depends on the position of the mountain as:
\begin{align}
    h_{0} \propto \cos^{2} \beta
\end{align}
This shows that the position of the mountain also plays a role in determining the amplitude of the GW signal, with maximum amplitude generated when the mountains are at the magnetic poles.

For \textit{model I}, we have $R(\beta) = R = R_{p} = R_{eq}$. As a result, the position of the mountains would not affect the amplitude of GW signals. This can also be understood because the \textit{model I} is symmetric, and one can always choose a coordinate system such that one of the axes passes through the mountain.

%----------------------------Mass of The Mountains -------------------------
\subsection{Effects of the mass of the mountains and deformation of the star}

The GW amplitude of the spherical accreting pulsars depends only on $\delta m$ eq \eqref{ho}. For the deformed star, both $\delta m$ and the deformation (the ratio of $\frac{R_p}{R_{eq}}$) are the determining factor eq \eqref{h0_modelII}. Table \ref{tab:my_label} shows these impacts on the GW amplitude. As expected, the GW of the spherical star only depends on $\delta m$ and decreases with a decrease in $\delta m$. 

In the case of a deformed star, the dependence is more complicated. For large deformation (comparatively small ratio of $\frac{R_p}{R_{eq}}$), the GW amplitude is determined mostly by the deformation. However, as the deformation decreases, the $\delta m$ value also becomes significant in determining the GW amplitude. 

 \begin{table}
     \centering
     \begin{tabular}{c c c c}
     \hline
          & \textbf{\textit{model I}} & \textbf{\textit{model II}} \\ \hline
          $\delta m$ [$M_{\odot}$]  & |$h_{0}$|  & $R_p/ R_{eq}$ &|$h_{0}$| \\  \\  \hline  \\
         $10^{-4}$ & $3.03 \times 10^{-23}$ &$0.80$ &   $1.91 \times 10^{-20}$  \\  
         &  &  $10^{-2}$& $6.41 \times 10^{-23}$\\ 
          &  &  $10^{-4}$& $3.31 \times 10^{-23}$\\ \hline \\
          $10^{-6}$& $3.03 \times 10^{-25}$ & $0.80$& $1.91 \times 10^{-20}$  \\  
          &  & $10^{-2}$& $9.72 \times 10^{-23}$  \\   &  &  $10^{-4}$& $6.47 \times 10^{-25}$\\\\  \hline \\ 
            $10^{-8}$& $3.03 \times 10^{-27}$ &  $0.80$ &$1.91 \times 10^{-20}$  \\
           &  &  $10^{-2}$& $9.75 \times 10^{-23}$\\ 
          &  &  $10^{-4}$ &$9.78 \times 10^{-25}$ \\ \\  \hline
     \end{tabular}
     \caption{Table showing the effects of the mass of the mountains on the amplitude of GW signal before starquake. For \textit{model II} we have changed the radius ratio along with the mass of the mountains.}
     \label{tab:my_label}
 \end{table}

%----------------------------Effects of value of sigma -------------------------
\subsection{Effects of \texorpdfstring{$\sigma$}{Lg} in the breaking frequency and the effects of distance of the source in GW signal}

The value of the breaking frequency depends on the value of $\sigma_{max}$. As there is no clear agreement on its value, we vary its value and check its dependence on the breaking frequency. Table \ref{tab:my_label2} shows that the breaking frequency increases with the increase in the value of $\sigma_{max}$. Hence, a suitable choice of $\sigma_{max}$ can be made to attain the breaking frequency at a significantly earlier time. 

\begin{table} 
     \centering
     \begin{tabular}{c c }
     \hline
          $\sigma_{max}$   & $\nu_{b}$ [Hz] \\  \hline  \\
         $10^{-1}$ & $918.2$  \\  \\ \hline \\
          $10^{-2}$& $290.4$  \\ \\  \hline \\
          $10^{-3}$& $91.2$  \\ \\  \hline
     \end{tabular}
     \caption{Table showing the dependence of the breaking frequency on the value of $\sigma_{max}$.}
     \label{tab:my_label2}
 \end{table}

The analysis of the two models was done by considering the source to be located at a distance of 10 kpc. The amplitude of the GW is inversely related to the distance of the source from the centre $h_{0} \propto \frac{1}{r}$ as seen in eq \eqref{ho}. On changing the distance of our source to 100 Mpc with the mass of the mountains $\delta m = 10^{-4} M_{\odot}$, the value of |$h_{0}$| for the \textit{model I} and \textit{model II} becomes of the order of $10^{-27}$ and $10^{-25}$ respectively. Thus, the location of the sources plays a major role in the detection of GW signals by the LIGO-VIRGO-KAGRA and future detectors.
%====================================Summary===========================================================
\section{Summary and Conclusion} \label{summary}

This work addresses the possibility of a change in continuous GW signal in recycled pulsars due to starquakes. We consider two models. In one of the models, we consider a spherically symmetric star consisting of mountains at the magnetic poles (\textit{model I}). In the second scenario, we consider a deformed star characterised by mountains (\textit{model II}) at its magnetic poles. The occurrence of mountains and deformation in the star's shape results in the emission of continuous GW. We consider that the star has a fluid core and an elastic crust. As the star spins up due to accretion, the crust experiences stress. Upon attaining the breaking frequency, it results in a starquake. The starquake suddenly changes the ellipticity, resulting in a sudden change in the continuous GW signal.\\

 \indent The analysis of the models interestingly shows clear, distinct signatures of the GW after the starquake. With accretion, the star spins up, and there is a gradual increase in the GW amplitude. Also, the strain on the stars' elastic crust increases as the star spins up. As it attains the breaking frequency, it suffers a starquake and, thereby, a sudden increase in the GW amplitude. 
 %Both these events lie under the current detector sensitivity range as shown in fig \ref{Starquake_Plots}. 
 The strength of the GW signal is parameter-dependent; however, if a starquake happens in a nearby NS, the change in the GW signal is well within the current detector sensitivity range. As the continuous GW emission suddenly changes, detectors can identify starquakes as a unique signature. 
The change in amplitude of GW before and after starquakes can also be understood physically. As the matter accumulates on the star's magnetic poles, the strain on its crust continues to build. When the breaking frequency is reached, the star undergoes a rapid shift in its shape, resulting in a starquake. This event causes a significant alteration in the star's quadrupole moment, which, in turn, significantly affects the GW amplitude. This shows that starquakes of spinning up recycled millisecond pulsars are characterised by a sudden increase in their continuous GW signal irrespective of their shape.
For a spherical star, the GW amplitude depends only on the accumulated accreted mass, whereas for a deformed star, it depends both on the accumulated mass and its deformation.

 \indent The prospect of detection of this sudden change in the continuous GW signal, as expected, depends on the distance of the location of the star. It also depends on the present rotational velocity of the recycled pulsars. If the star's present rotational velocity is quite different from its breaking frequency (for which the starquake would happen), then it would take an enormous amount of time to reach its breaking frequency. However, if its rotational velocity is near the breaking frequency, the change in the continuous GW signal is most likely to be observed with dedicated detectors. Usually, the breaking frequency for a star lies in the range of a few hundred Hz (it depends on the star mass) \citep{fattoyev2018}, and recycled millisecond pulsars are also seen to have a frequency of a few hundred Hz. Presently, the detected millisecond pulsar count is more than a hundred, and as they rotate very fast, there is a high chance that a few of them can undergo starquakes. If we can detect continuous GW signals, observing the recycled millisecond pulsars for a substantially long time can result in the detection of starquakes in them.

 \indent It is to be understood that the calculation done in this work is for two simple models. Although realistic models could be more complicated, this analysis provides an overall picture of the accreting millisecond pulsars. There are a few areas in which the calculation could be improved, one of them being the explicit dependence of stress on the accreting matter. Although the source of the stress is an infalling matter, it is not taken explicitly in the calculation. It is avoided by assuming the difference in the crust and core rotational velocity (which appears because of the infalling matter). However, to have a more accurate model, the infalling matter's accretion rate and mass should be considered. In future work, we plan to simulate the dynamic evolution of the stars under the effect of accretion and the triggering of starquakes.
 
 %========================Acknowledgement=====================================
\section*{Acknowledgment}
The authors would also like to thank IISER Bhopal for providing the infrastructure for this work. The authors also acknowledge the helpful discussions of Pallavi Bhat, Debojoti Kuzur, Shamim Haque, and Shailendra Singh. The authors thank Garvin Yim for carefully reading the manuscript and suggesting modifications. SC would also like to acknowledge Manoel Felipe Sousa and Elia Giliberti for their discussions and assistance in this project. SC acknowledges the Prime Minister's Research Fellowship (PMRF), Ministry of Education Govt. of India, for a graduate fellowship. RM acknowledges the Science and Engineering Research Board (SERB), Govt. of India, for monetary support in the form of a Core Research Grant (CRG/2022/000663). KKN would like to acknowledge the Department of Atomic Energy (DAE), Govt. of India, for sponsoring the fellowship covered under the sub-project no. RIN4001-SPS (Basic research in Physical Sciences).
%%%%%%%%%%%%%%%%%%%%%%%%%%%%%%%%%%%%%%%%%%%%%%%%%%
\section*{Data Availability}

This is a theoretical work; hence, it does not have any additional data.


% The best way to enter references is to use BibTeX:\Sc

\bibliographystyle{mnras}
\bibliography{accretion} % if your bibtex file is called example.bib

\appendix

%=======================================================================================
%==================================FLE Model===========================================
%=======================================================================================
\section{FLE Model} \label{FLE}

FLE model (\cite{Franco, Giliberti}) inside the star describes the core as fluid and the crust as elastic, having the same densities. This section shows how we calculated the displacement field $\textbf{u}$.  With time, due to accretion, there is a difference in the rotational velocity of the fluid core and the outer elastic but brittle envelope. To consider this, we assume two stars: one with rotational velocity $\Omega$ (the outer envelope) and another with rotational velocity $\Omega$ - d $\Omega$ (the inner fluid core). The net displacement between the two configurations can then be approximated as :
\begin{equation} \label{disp}
    \textbf{u} = \textbf{u}_{\Omega } - \textbf{u}_{\Omega -d\Omega} \propto d\Omega \: \Omega
\end{equation}
This gives the value of \textbf{u} as:
\begin{equation} \label{ur}
       u_{r}(r,\theta) = (ar - \frac{1}{7}Ar^{3} - \frac{1}{2} \frac{B}{r^{2}} + \frac{b}{r^{4}})P_{2} 
\end{equation}
\begin{equation} \label{ut}
    u_{\theta}(r,\theta) = \left(\frac{1}{2}ar - \frac{5}{42}Ar^{3} - \frac{1}{3} \frac{b}{r^{4}}\right)\frac{dP_{2}}{d \theta}
\end{equation}
where $P_{2} = \frac{1}{2}(3 \cos ^{2} \theta - 1)$ is the second Legendre polynomial. The four coefficients $a,b,A,B$ are found using four boundary conditions, two at the crust-core transition radius and two at the star's surface. Using the boundary conditions eq \eqref{ur} and eq \eqref{ut} can be rewritten as :
\begin{equation} \label{urr}
    u_{r} = \dfrac{\Omega \: d \Omega \: R^{3}}{Q(c_{t},\nu_{k},L)} \left(\frac{\Tilde{a} r}{R} - \frac{\Tilde{A} r^{3}}{7 R^{3}} - \frac{\Tilde{B} R^{2}}{2r^{2}} + \frac{\Tilde{b}R^{4}}{r^{4}}\right) P_{2}
\end{equation}
\begin{equation} \label{utt}
    u_{\theta} = \dfrac{\Omega \: d \Omega \: R^{3}}{Q(c_{t},\nu_{k},L)} \left(\frac{\Tilde{a} r}{2R} - \frac{5\Tilde{A} r^{3}}{42 R^{3}} - \frac{\Tilde{b}R^{4}}{3r^{4}}\right) \frac{dP_{2}}{d\theta}
\end{equation}
Where $\Tilde{a},\Tilde{b},\Tilde{A},\Tilde{B},$ are dimensionless and depends on $L = R'/R$. $R'$ denotes the radius of the core, and $R$ denotes the star's radius. The transverse speed of sound in the crust is given by $c_{t}$ and $\nu_{k}$ denotes the Keplerian velocity ($v_k  = \sqrt{G M / R}$, M is the mass of the star and R the radius). The values of the coefficients in eq \eqref{urr} and eq \eqref{utt} can be given as :
\begin{equation}
    \Tilde{a} = 280 (13q^{2}-7q+2)
\end{equation}
\begin{equation}
    \Tilde{b} = -5(643q^{2}-232q+37)
\end{equation}
\begin{equation}
    \Tilde{A} = 280(15q^{2} - 13q + 5)
\end{equation}
\begin{equation}
    \Tilde{B} = -560(70q^{2} - 27q + 5)/3
\end{equation}
\begin{equation}
    Q/ \nu_{k}^{2} = 35(q^{2}(109-240 \chi^{2})-q(48 - 60 \chi^{2}) + 11)
\end{equation}
where $q=1-L$ and $\chi = \frac{c_{t}}{\nu_{k}} \ll 1$.

It is important to note that the calculation of the deformation coefficients is provided in the cited works and presented here for the reader's clarity. More detailed calculations can be found in the referred calculations.

% Don't change these lines
\bsp	% typesetting comment
\label{lastpage}
\end{document}